\documentclass[journal,twoside,web]{ieeecolor}
\usepackage{generic}
\usepackage{cite}
\usepackage{amsmath,amssymb,amsfonts}
\usepackage{algorithmic}
\usepackage{graphicx}
\usepackage{algorithm,algorithmic}
\usepackage{hyperref}
\hypersetup{hidelinks=true}
\usepackage{textcomp}
\usepackage{array}
\usepackage{multirow}
\def\BibTeX{{\rm B\kern-.05em{\sc i\kern-.025em b}\kern-.08em
    T\kern-.1667em\lower.7ex\hbox{E}\kern-.125emX}}
\begin{document}
\title{A Machine-to-Machine Knowledge-Guided LLM Agent for Generalizable Radiotherapy Treatment Planning}
\author{Md Mainul Abrar, Xun Jia, Yujie Chi
\thanks{This work was supported in part by the National Institutes of Health (NIH) under grants R21EB036619 and R01CA254377.}
\thanks{Md Mainul Abrar (e-mail: mxa0419@mavs.uta.edu) and Yujie Chi (e-mail: yujie.chi@uta.edu) are with the Department of Physics, The University of Texas at Arlington, Arlington, 76019, TX}
\thanks{Xun Jia (e-mail: xunjia@jhu.edu) is with the Department of Radiation Oncology and Molecular Radiation Sciences, Johns Hopkins University, Baltimore, MD}
\thanks{English writing of this work has been polished under the aid of OpenAI GPT.}
}

\maketitle

\begin{abstract}
In this work, we propose a prototype machine-to-machine (M2M) knowledge-guided Large Language Model (LLM) framework for automated radiotherapy treatment planning. In the proposed paradigm, Treatment Planning Parameter (TPP) distribution knowledge discovered by a Deep Reinforcement Learning (DRL) agent is transferred to an LLM agent through in-context learning, enabling autonomous iterative planning without human intervention. While standard LLM-based planning often lacks physical intuition and struggles with convergence, the integration of DRL-derived guidance constrains the agent to a physically valid parameter space. Experimental evaluations are performed across three diverse planning scenarios: basic prostate cases, complex prostate configurations with increased organ-at-risk (OAR) constraints, and liver cases. The evaluation results demonstrate that the guided LLM agent consistently achieves optimal planning scores while significantly reducing the number of iterations compared to unguided planning. Analysis of the final TPP configurations reveals that the agent successfully learns a hierarchical priority of objectives, effectively restoring a logical "cause-and-effect" relationship between parameter tuning and dosimetric outcomes. Crucially, this prototype framework exhibits robust generalizability, maintaining high planning quality regardless of specific patient anatomy, treatment site, or initial plan quality. By bridging the specialized optimization of DRL with the adaptive reasoning of LLMs, this M2M framework establishes a scalable foundation towards generalizable autonomous treatment planning, ultimately benefiting clinical practice in realistic environments.
\end{abstract}


\section{Introduction}
\label{sec:introduction}
\IEEEPARstart {A} central goal of cancer radiotherapy is to maximize tumor control while minimizing radiation-related toxicities to surrounding organs at risk (OARs). In modern radiation clinics, this is typically achieved through complex beam modulation techniques, such as intensity-modulated radiation therapy (IMRT) utilizing multi-leaf collimators \cite{intensity2001intensity, teoh2011volumetric}. Although modulating hundreds of beamlets across multiple angles enables highly conformal dose distributions, it introduces the high-dimensional challenge of identifying the optimal beam configuration for each patient. In practice, this process is framed as a two-level inverse optimization problem within a treatment planning system (TPS). At the lower level, a numerical optimizer determines the machine parameters ($x$), specifically the high-dimensional beamlet intensity vector, that best satisfy a defined objective function. However, the performance of this optimizer is fundamentally governed by the upper-level configuration of treatment planning parameters (TPPs), such as weighting factors ($\lambda$), dose thresholds ($t$), and volume constraints ($V$). Because the relationship between a specific set of TPPs and the resulting plan quality is complex and non-linear, a manual trial-and-error navigation of the TPP space is often needed to reach a clinically acceptable solution. Consequently, while the determination of beam intensities ($x$) is automated, the upper-level tuning of TPPs still relies on labor-intensive manual intuition. This often leads to suboptimal plans under clinical time constraints and significantly limits the scalability of modern radiotherapy facilities\cite{hansen2022plan}.

Various strategies have been developed to mitigate this "human-in-the-loop" bottleneck. Early efforts primarily include Knowledge-Based Planning (KBP) \cite{ge2019knowledge, babier2021openkbp, momin2021knowledge, robinson2023can, wang2026universal} and Multi-Criteria Optimization (MCO) \cite{kufer2003intensity, monz2008pareto, breedveld2019multi}. These approaches focus on providing a superior starting point for the human planner by predicting achievable dose objectives or generating a Pareto-optimal trade-off surface. However, a fundamental "translation gap" remains: these methods do not provide a direct mechanism to map high-level dose predictions into the actionable TPP configurations required to realize the intended plan quality. Consequently, the final deliverable plan still heavily relies on the planner's individual expertise. Furthermore, KBP is inherently data-driven and site-limited , while MCO often relies on convex approximations that may not guarantee the achievability of predicted objectives within the actual non-convex planning environment. More recently, research has shifted toward replacing the manual upper-level loop with autonomous agents based on Deep Reinforcement Learning (DRL) \cite{mnih2015human, silver2017mastering} and Large Language Models (LLMs) \cite{vaswani2017attention}. While promising, DRL agents are often restricted by rigid reward function designs and specific input configurations, which limits their ability to adapt to changes in dose objectives or anatomical sites\cite{shen2019intelligent, shen2020operating, shen2021improving, shen2021hierarchical, gao2023implementation, sprouts2022development, abrar2025actor, yang2025automated}. Conversely, while LLM-based planners offer high flexibility for diverse planning task configurations, they frequently lack domain-specific optimization knowledge and robust reasoning capabilities. Consequently, these models often rely on human-derived TPP tuning principles or interactive conversational guidance to navigate the parameter space\cite{wang2025feasibility, liu2025automated, wei2025feasibility, yang2025zero, jafar2026human}, failing to achieve cross-domain generalizability. 

We argue that existing approaches remain fragmented and site-specific primarily due to the lack of systematic investigation into the universal principles governing TPP interactions. In this work, we address this critical gap by proposing a fundamental paradigm shift: formulating TPP configuration as a generalizable control problem. This approach is motivated by the observation that the TPS utilizes an identical two-level optimization engine across diverse tumor sites and clinical protocols. Furthermore, our prior research revealed that a DRL agent trained on a single patient case can achieve high-quality treatment planning for prostate IMRT cases across diverse anatomies and beam configurations, and the resulting TPP configurations exhibited highly patterned behavior \cite{abrar2025actor, mainul2025new}. Together, these findings suggest that there exists a globally workable structure for the upper-level TPP-tuning configurations and that DRL’s in-depth exploration power can help elucidate this structure. 

In this work, we introduce a machine-to-machine (M2M) knowledge-guided LLM planning framework designed to achieve generalizable automatic treatment planning. The key idea is to utilize the exploratory power of a DRL agent to identify universal TPP distributions and enable an LLM agent to effectively apply this information for diverse planning scenarios. Specifically, building on the cross-case generality demonstrated by the DRL agent in our prior work, we extracted TPP configurations from successful prostate IMRT plans (one target, two OARs, 7-beam configuration) \cite{abrar2025actor, mainul2025new} and integrated them as domain knowledge through in-context learning (ICL) \cite{brown2020language} to guide the LLM-based planning process. We evaluated the framework across three scenarios of increasing complexity: (1) prostate IMRT under the original 7-beam configuration, (2) prostate IMRT with four OARs under a 180-beam configuration, and (3) liver IMRT with two OARs under a 7-beam configuration. Our key findings and contributions are as follows:
\begin{itemize}
    \item The DRL-guided LLM agent consistently achieved full-score planning across all evaluated scenarios, significantly outperforming unguided LLM planning in both target coverage and OAR sparing.
    
    \item The framework demonstrated superior efficiency, reaching optimal plans within 2–6 iterations, whereas the unguided version required 12–20 steps and frequently failed to meet plan evaluation criteria.

    \item TPP-tuning knowledge is highly transferable: strategies derived from 7-beam prostate cases successfully automated planning for complex 180-beam prostate and 7-beam liver cases without any site-specific retraining.

    \item Quantitative analysis demonstrated that consistent TPP distribution patterns yield high-quality plans across diverse scenarios, confirming a site-invariant tuning structure that maintains a stable cause-and-effect relationship within the optimization space.
\end{itemize}

\section{Related Work}
\subsection{Traditional Automation: KBP and MCO}
KBP approaches \cite{ge2019knowledge, babier2021openkbp, momin2021knowledge, robinson2023can, wang2026universal} utilize statistical models or deep learning to predict achievable dose-volume histograms (DVH) or voxel-level dose distributions based on historical treatment plans. While early KBP models were predominantly site-specific, recent progress has aimed at achieving cross-site generality. For instance, UniDose utilizes generalized input channels to accommodate a wide range of disease sites and arbitrary beam configurations within a single model \cite{wang2026universal}. However, the model is still data-driven, the prediction capability of which is still restricted by the quality and complexity of the training data.

MCO\cite{kufer2003intensity, monz2008pareto, breedveld2019multi}, on the other hand, generates a set of Pareto-optimal plans, allowing planners to navigate trade-offs between competing clinical objectives. However, MCO typically relies on convex approximations of the planning space, which may not guarantee that the selected trade-off is mathematically achievable within the non-convex environment of modern TPS. 

More critically, both KBP and MCO serve as decision-support tools rather than fully autonomous systems, as they still require human experts to manually bridge the gap between predicted dose objectives and TPP configuration to get the final deliverable treatment plan. 

\subsection{Autonomous Agents: DRL and LLMs}
In DRL, an agent is trained to make sequential decisions in a dynamic environment to maximize cumulative rewards via full self-exploration, and has demonstrated superhuman performance in domains such as Atari games and the game of Go \cite{mnih2015human, silver2017mastering}. Motivated by these successes, our group and others have explored DRL-based automatic TPP adjustments to dynamically optimize a treatment plan until satisfying evaluation criteria \cite{shen2019intelligent, shen2020operating, shen2021improving, shen2021hierarchical, gao2023implementation, sprouts2022development, abrar2025actor, yang2025automated}. In particular, Abrar et. al. demonstrated that a well-trained DRL agent can achieve robustness and generalization across patient cases with varying anatomical structures and beam configurations within a same treatment site \cite{abrar2025actor}. Moreover, DRL agents can implicitly learn clinically meaningful decision strategies, such as identifying dose violation regions from input data and selecting appropriate TPP adjustments to mitigate them, even though these violations are encoded only in the reward function and are not explicitly provided as inputs \cite{mainul2025new}. Despite these successes, DRL agents are fundamentally constrained by the rigid architecture of their reward functions and input configurations, which limits their ability to generalize across diverse treatment sites or varying numbers of OARs without extensive re-training.

To address this lack of flexibility, recent research has explored LLMs for radiotherapy planning. Unlike the rigid input-output mappings of DRL, LLMs' transformer architecture with self-attention mechanisms \cite{vaswani2017attention} offers high adaptability to diverse clinical instructions and variable objective sets. However, general-purpose LLMs lack the deep, domain-specific optimization logic required to navigate the non-linear TPP space. As a result, current LLM-based planners often rely on human-centric guidance, including historical TPP configurations from successful clinical plans \cite{wang2025feasibility, liu2025automated, wei2025feasibility}, summarized expert planning principles and heuristic parameter ranges \cite{yang2025zero}, or real-time interactive conversational steering \cite{jafar2026human}. While these studies show effectiveness, these approaches are site-specific and the planning quality and scope is highly restricted by human's intuition on TPP tunings. 

\section{Methods}
\begin{figure*}[hbt!]
    \centering
    \includegraphics[width=1\linewidth]{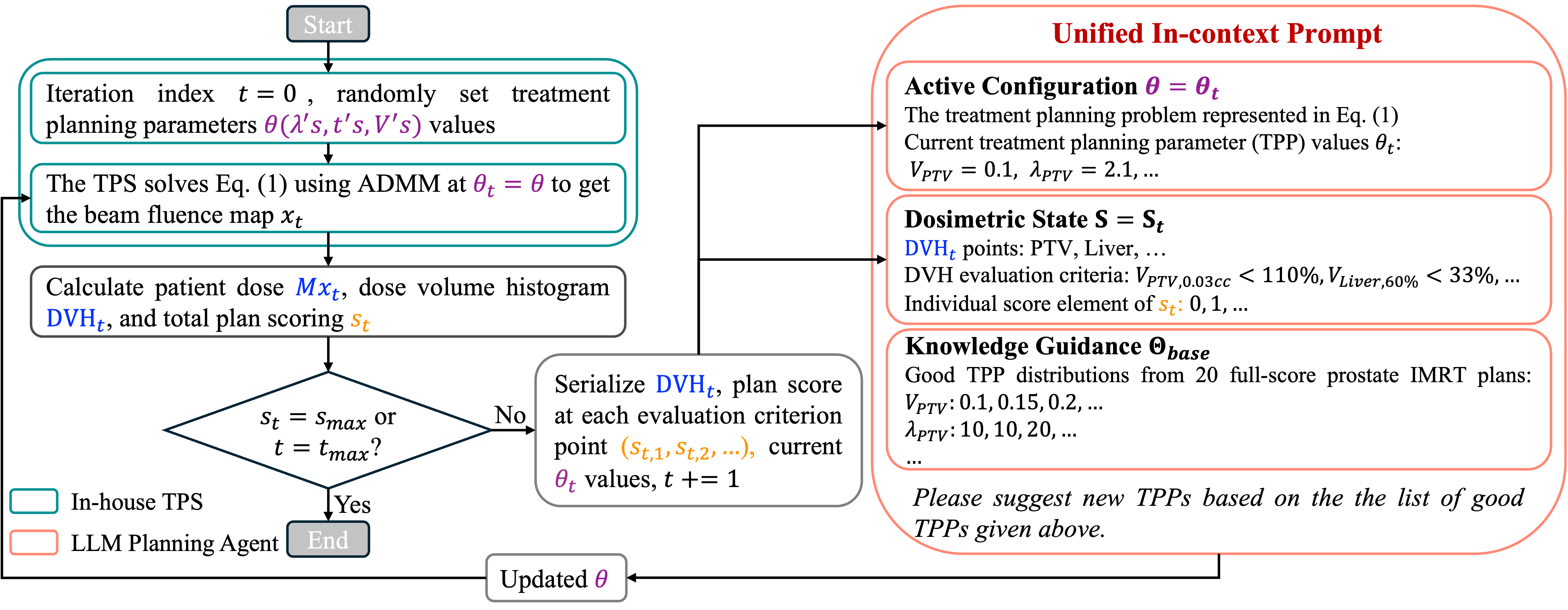}
    \caption{Schematic illustration of the proposed machine-to-machine (M2M) knowledge-guided large language model (LLM)-based autonomous treatment planning framework.}
    \label{fig:overallWorkflow}
\end{figure*}

In this section, we introduce the M2M LLM-guided planning framework, an autonomous, closed-loop control system designed to bridge the gap between high-level clinical objectives and low-level TPP parameter configuration. The overall workflow is illustrated in Fig. \ref{fig:overallWorkflow}. The architecture integrates an LLM-based reasoning agent with an in-house TPS. The optimization process follows an iterative bi-level logic. It initiates with an arbitrary TPP configuration $\theta_0$, after which the TPS executes a dose-volume-constrained optimization. The resulting dosimetric state and evaluation result are then serialized and synthesized with DRL-derived knowledge in the LLM agent to predict an updated $\theta'$ for the next iteration. This iterative cycle continues until all clinical criteria are satisfied or the maximum tuning budget is exhausted. The following sections describe each component of the framework in detail.
\subsection{The In house Inverse Treatment Planning System}\label{sec:TPS}
To evaluate the framework, we utilized an in-house dose-volume-constrained TPS \cite{sprouts2022development, abrar2025actor} that adheres to Varian Eclipse functional specifications \cite{varian2015eclipse}. The lower-level optimization is formulated as a constrained minimization of a quadratic objective function:

\begin{multline}
    min \frac{1}{2}||Mx-d_p||^{2}_{-}+ \frac{\lambda}{2}||(Mx-td_p)_{V_{\mathrm{PTV}}}||^{2}_{+}+ \\ \sum_i^N \frac{\lambda_i}{2}||(M_ix-t_{i}d_p)_{V_i}||^{2}_{+}, \\
    \textrm{s.t. } x \geq 0, D_{95\%}(Mx) = d_p.
\end{multline}

Here,  $|\cdot|^2_-$ and $|\cdot|^2_+$ denote one-sided quadratic penalties applied to negative and positive components, corresponding to under-dose and over-dose violations, respectively. Adequate target coverage is strictly enforced via a $D_{95\%}$ hard constraint. The TPP vector $\theta$, containing weights $\lambda$, volume fractions $V$, and dose thresholds $t$ for both PTV and OARs ($i=1, ..., N$), defines the search space for the upper-level agent. For any given $\theta$, the resulting non-convex problem is solved using the Alternating Direction Method of Multipliers (ADMM), ensuring consistent numerical convergence across all sites.

\subsection{M2M Knowledge Extraction and Representation}\label{sec:TPPSpace}
The core of our M2M approach lies in the extraction of latent TPP optimization logic from a high-performance DRL agent previously trained for prostate IMRT \cite{abrar2025actor}. In that work, despite being trained on a single patient case, the agent demonstrated robust generalization by producing full-score plans for diverse test cases with significantly varied patient anatomies and beam configurations. Attribution analysis via Integrated Gradients (IG) revealed that the agent effectively learned to identify specific dose-violation regions from DVH inputs and developed a global tuning strategy to promote actions that mitigate these violations across multiple structures simultaneously \cite{mainul2025new}. This systematic strategy is evidenced through a decisive policy distribution characterized by low entropy and high stability at each individual tuning step, which consistently converges toward a highly patterned and localized final TPP distribution.

While our previous work examined the full patient dataset to identify these TPP patterns, this study utilizes a subset of 20 successful planning instances to construct the knowledge base, reserving the remaining cases for independent validation. These 20 patient cases encompass diverse anatomies with a standard configuration of one PTV and two OARs (bladder and rectum) under a 7-beam arrangement with fixed gantry angles at $0^\circ$, $32^\circ$, $64^\circ$, $96^\circ$, $264^\circ$, $296^\circ$, and $328^\circ$. 

\begin{figure}[hbt!]
    \centering
    \includegraphics[width=0.9\linewidth]{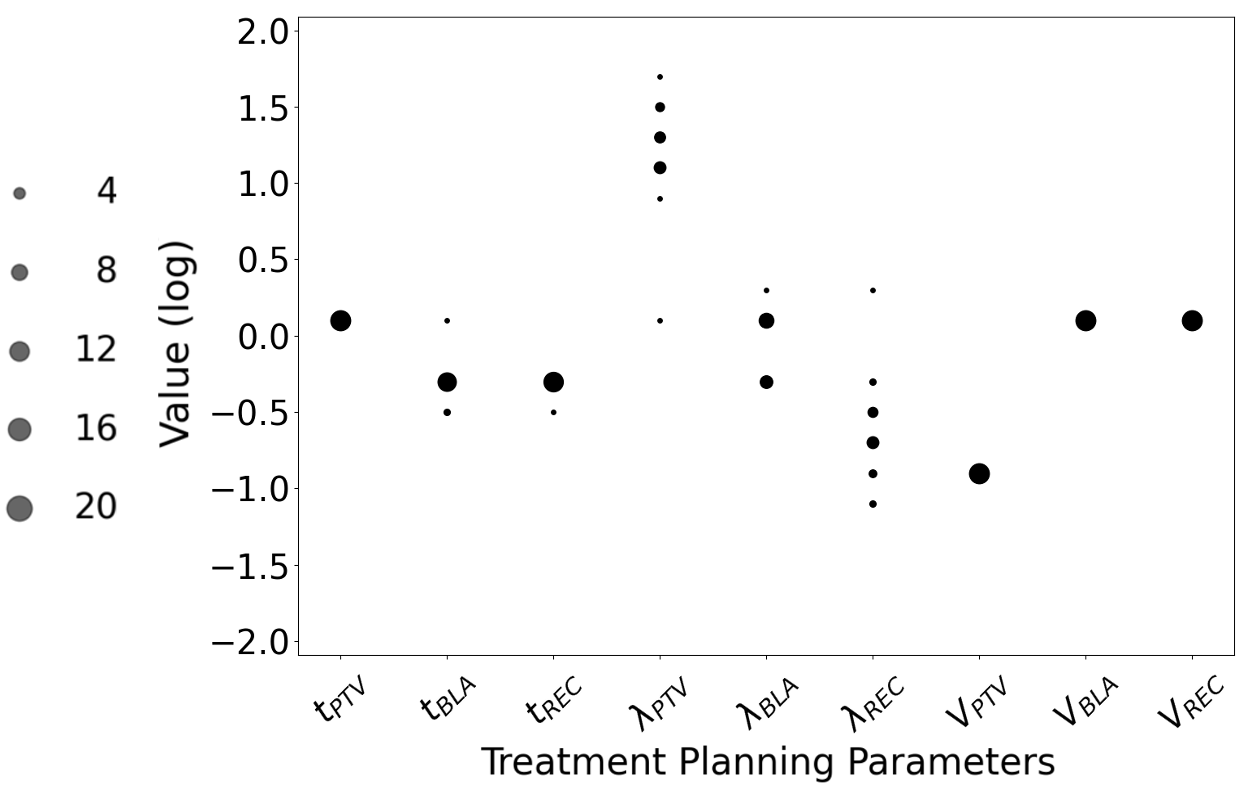}
    \caption{The treatment planning parameter (TPP) space for successful planning generated by the ACER agent \cite{abrar2025actor}, extracted from 20 prostate IMRT patient cases with one target and two organs-at-risk (OARs).}
    \label{fig:TPPSpaceGuidance}
\end{figure}

The corresponding TPP distributions are illustrated in Fig. \ref{fig:TPPSpaceGuidance}. As shown, parameters associated with target coverage ($t_{PTV}$, $V_{PTV}$) and OAR volume fractions ($V_{BLA}$, $V_{REC}$) exhibit near-singular convergence, while penalty weights ($\lambda$) and thresholds ($t$) cluster within well-defined, preferential ranges. The emergence of this structural invariance, where optimal parameter settings remain consistent regardless of anatomical complexity, suggests the existence of a universal optimization manifold within the TPP space and serves as the theoretical foundation for the M2M framework. By serializing this patterned manifold into a structured knowledge base, the LLM agent performs an optimization policy transfer, inheriting the DRL agent's ability to distinguish the relative importance of tuning actions without requiring human-derived heuristics.

\subsection{The LLM Planning Agent}
We utilized GPT-4o \cite{hurst2024gpt} as the high-level reasoning engine within the bi-level control architecture to oversee iterative TPP adjustments. The agent's decision-making process is performed through few-shot ICL, a paradigm that enables the model to perform complex tasks at inference time by conditioning on a prompt containing a set of demonstrations. This mechanism allows the LLM to adapt the DRL agent's TPP tuning logic to new anatomical scenarios without requiring additional weight updates of the LLM framework or extensive retraining of the source DRL agent. 

At each iteration $t$, the state-space is serialized into a natural language prompt comprising three functional components:
\begin{enumerate}
    \item \textbf{Active Configuration ($\theta_t$)}: The inverse optimization problem itself and the current numerical state of the TPP vector ($\lambda, t, V$) utilized by the lower-level optimizer.
    \item \textbf{Dosimetric State ($S_t$)}: Uniformly sampled DVH points on each DVH curve, specific clinical plan checkpoint and corresponding violations identified from the current treatment plan.
    \item \textbf{Knowledge Guidance ($\Theta_{base}$)}: A serialized collection of the 20 final values for each TPP variable from the planning cases described in Section \ref{sec:TPPSpace}. To ensure LLM's decision is based on the M2M policy transfer, we explicitly direct the agent to 
    \end{enumerate}
    
It's worth noting that to ensure an objective M2M knowledge transfer, we performed three complementary architectural designs: 1) to prevent the LLM agent from inheriting inter-parameter correlations tied to the specific anatomical geometries of the source prostate cases, we adopt a parameter-wise serialization strategy, representing $\Theta_{base}$ as independent, parameter-wise distributions rather than in a patient-list format that would implicitly couple TPP vectors with individual anatomies; 2) the LLM agent is explicitly instructed to condition its $\theta'$ outputs on the provided knowledge base, ensuring that the generation process is driven by the encoded TPP distributions rather than the model’s internal priors or built-in reasoning patterns; and 3) no human-based interpretation, intervention, or heuristic guidance is introduced, thereby enforcing a fully M2M paradigm and eliminating potential sources of human-induced bias.

By synthesizing current failure modes ($S_t$) with successful TPP distribution patterns, the LLM predicts an updated TPP set $\theta'$. This process is continued until a full-score plan is obtained or the iterative planning reaches a maximum of 20 steps.

\subsection{Datasets} \label{sec:data}
We performed a systematic evaluation of the proposed framework under three scenarios: (1) 10 prostate IMRT cases with one PTV and two OARs (bladder and rectum) under a 7-beam configuration; (2) 10 prostate IMRT cases with one PTV and four OARs (bladder, rectum, and left and right femoral heads) under a 180-beam configuration; and (3) 10 liver IMRT cases generated using random TPP initializations (thus different starting dosimetric states) from a single patient case under a 7-beam configuration. This scenario represents a rigorous out-of-distribution (OOD) test for the stability and robustness of the prostate-derived TPP-tuning guidance.

The prostate cases in scenarios (1) and (2) were drawn from the dataset described in our previous work \cite{mainul2025new}, but are independent from the 20 cases used to construct the TPP guidance in Subsection \ref{sec:TPPSpace}. The beam angles in scenario 1 are identical to those used in the TPP space guidance cases. In scenario 2, beam angles are evenly distributed over a full circle with a $2^\circ$ interval. The liver case in scenario 3 was obtained from the public Common Optimization for Radiation Therapy (CORT) dataset \cite{craft2014shared}, from which the beam configuration $(58^\circ, 106^\circ, 212^\circ, 216^\circ, 226^\circ, 296^\circ, 328^\circ)$ with couch angles $(0^\circ, 0^\circ, 0^\circ, 32^\circ, -13^\circ, 17^\circ, 0^\circ)$ was selected.

\subsection{Evaluation Metrics and Experimental Design}
Treatment plans were evaluated using the dose–volume criteria summarized in Tables \ref{tab:prostate_criteria} and \ref{tab:liver_criteria}. These metrics follow the ProKnow scoring system for prostate cancer (ProKnow Systems, Sanford, FL, USA) and the Dose–Volume Constraints for Organs at Risk in Radiotherapy (CORSAIR) guidelines for liver radiotherapy \cite{bisello2022dose}. Plans were scored in binary (1 for satisfaction, 0 for violation) per criterion, resulting in maximum achievable scores of 9, 15, and 4 for the three respective scenarios.

To evaluate the impact of the M2M policy transfer, the LLM agent's performance was compared in two configurations:
\begin{enumerate}
    \item \textbf{Knowledge-Guided (Few-shot)}: The agent utilized the case-decoupled $\Theta_{base}$ distribution as in-context demonstrations to guide its reasoning.
    \item \textbf{Baseline (Zero-shot)}: The agent operated via identical instructions and state-space inputs but without the TPP guidance distributions, relying solely on its internal pre-trained logic.
\end{enumerate} 

Planning effectiveness and efficiency were quantified by the final plan score and the number of tuning steps required for convergence. Furthermore, to demonstrate how the universal TPP distributions inherited from the DRL agent facilitates superior generalization and planning stability, a multi-faceted statistical analysis was performed across all three testing scenarios. First, we analyzed the mean and standard deviation of organ volumes at specified dose-volume evaluation criteria to quantify plan quality beyond the binary scoring system. Second, the final TPP configurations were compared between the unguided and guided groups to identify differences in parameter search behavior. Finally, we evaluated the Spearman correlation coefficients between the final TPP values and organ dose-volume distributions to further quantify the degree of planning generality of the M2M framework.

\begin{table*}[hbt!]
\centering
\caption{Dose–volume plan evaluation criteria for prostate cancer IMRT used in test scenarios 1 and 2, with a prescription dose of 79.5 Gy to the PTV.}
\label{tab:prostate_criteria}
\resizebox{\linewidth}{!}{%
\begin{tabular}{c|c c c c|c c c c|c c c}
\hline
\multicolumn{1}{c|}{PTV} &
\multicolumn{4}{c|}{Bladder} &
\multicolumn{4}{c|}{Rectum} &
\multicolumn{3}{c}{Femoral Head} 
\\
\hline
$D_{0.03\,\mathrm{cc}}$ & $V_{81.2\%}$ &$V_{89.1\%}$ &$V_{94.3\%}$ &$V_{100.6\%}$ &   
$V_{75.5\%}$ &$V_{81.2\%}$ &$V_{89.1\%}$ &$V_{94.3\%}$  &   
$V_{37.7\%}$ &$V_{50.3\%}$ & $V_{55.3\%}$ 
\\
$<110\%$ &$<55\%$ &$<40\%$ &$<30\%$ & $<20\%$ & 
$<55\%$ &$<40\%$ & $<30\%$ & $<20\%$ & 
$<50\%$  & $<35\%$ &  $<5\%$
\\
\hline
\end{tabular}%
}
\end{table*}

\begin{table}[hbt!]
\centering
\caption{Dose–volume plan evaluation criteria for liver cancer IMRT used in test scenario 3, with a prescription dose of 50 Gy to the PTV.}
\label{tab:liver_criteria}
\begin{tabular}{c|c c|c}
\hline
\multicolumn{1}{c|}{PTV}& 
\multicolumn{2}{c|}{Liver} &
\multicolumn{1}{c}{Heart} 
\\
\hline
$V_{0.03\,\mathrm{cc}}$& $D_{\mathrm{Mean}}$ & $V_{60\%}$ & $V_{50\%}$  \\
$<110\%$& $<25$ Gy & $<33\%$ & $<10\%$  \\
\hline
\end{tabular}
\end{table}


\section{Results} \label{sec: results}
\subsection{LLM Planning Efficiency and Effectiveness}
\begin{figure*}[hbt!]
    \centering
    \includegraphics[width=\linewidth]{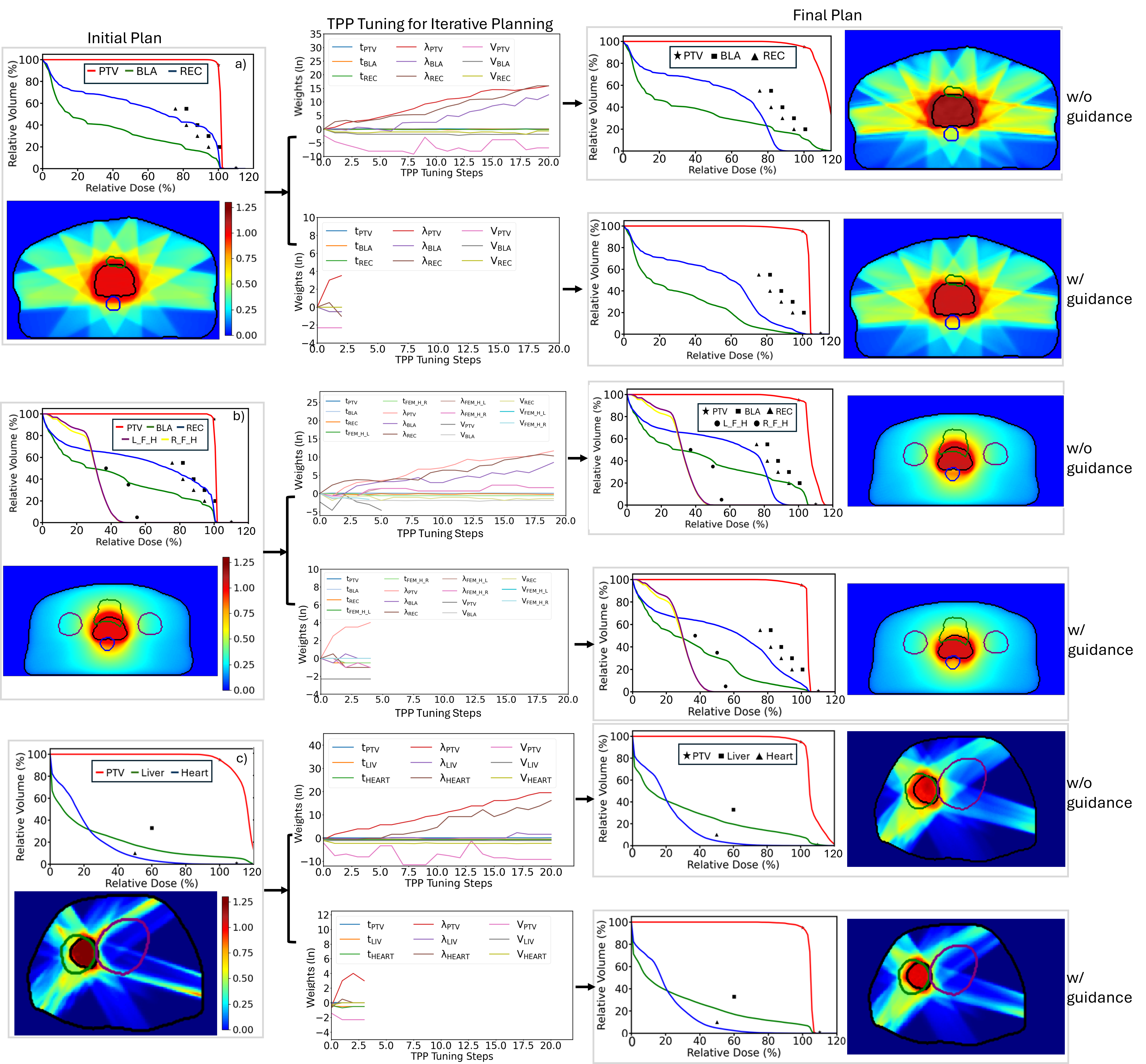}
    \caption{Representative planning cases illustrating the iterative treatment planning process performed by the LLM agent for planning scenarios (a) 1 (prostate IMRT with two OARs and a 7-beam configuration), (b) 2 (prostate IMRT with four OARs and a 180-beam configuration), and (c) 3 (liver IMRT with two OARs and a 7-beam configuration). For each scenario, results without (upper row) and with (lower row) DRL-derived TPP distribution guidance are shown. The star, square, triangle, and circle markers in the DVH plots indicate the dose evaluation criteria for the respective organs.}
    \label{fig:PlanningCasesAll}
\end{figure*}

Representative planning cases for the three test scenarios with and without guidance are illustrated in Fig. \ref{fig:PlanningCasesAll}. Although the two agents begin with the same initial plan, their tuning trajectories diverge significantly. The unguided agent makes aggressive, widespread adjustments across multiple TPP parameters but fails to meet all plan evaluation criteria. For instance, in scenario 1, it continuously increases all $\lambda$ values to an order as high as $e^{15}$ and suppresses $V_{\mathrm{PTV}}$ to almost 0 over 20 planning iterations without successfully regulating PTV hot spots. Conversely, the guided agent performs highly targeted TPP tuning and often achieves a full-score plan within a few steps. Still taking scenario 1 as an example, the guided agent converges in just two planning iterations, primarily by increasing $\lambda_{\mathrm{PTV}}$ to the order of $e^{4}$ and slightly reducing $\lambda$ values for the OARs.

{}

Statistical results for both planning scores and numbers of planning steps for the three scenarios are shown in Table \ref{tab:planningPerformance}. As is observed, the guided LLM agent achieves significantly higher final plan scores than the unguided version. In scenario 1, starting from an initial score of $6.0 \pm 1.8$, final planning score reaches $7.0 \pm 1.7$ without guidance and a maximum of $9.0$ with guidance, respectively. A similar pattern is observed in scenario 2. The final scores are improved to $13.6 \pm 2.0$ and $15.0$ from an initial score of $11.4 \pm 1.4$, respectively. In scenario 3, the score decreases from an initial value of $2.5 \pm 0.4$ to $2.2 \pm 0.4$ without guidance, while the guided approach again achieves the maximum score of $4.0$. In terms of planning efficiency, the guided agent spends significantly fewer steps in planning than the unguided version. It costs the unguided agent $18.1 \pm 4.1$, $12.4 \pm 6.5$, and $20.0 \pm 0.0$ steps for planning scenarios 1–3, respectively. In particular, the agent struggled to optimize scenario 3 (liver IMRT planning), where it exhausts the maximum 20 planning steps without achieving full-score plans ($2.2 \pm 0.4$ out of $4.0$). In contrast, the guided agent requires only $2.1 \pm 0.3$, $3.9 \pm 2.0$, and $5.3 \pm 2.9$ steps on average to achieve full-score plans.

\begin{table*}[hbt!]
\centering
\caption{Planning performance comparison across the three test scenarios with and without DRL-derived TPP space guidance, measured by the final plan score and the number of planning steps.}
\label{tab:planningPerformance}
\begin{tabular}{c c c cc cc c}
\hline
\multirow{2}{*}{Scenarios} &
\multirow{2}{*}{Planning Cases} &
\multirow{2}{*}{Initial Score} &
\multicolumn{2}{c}{Final Score} &
\multicolumn{2}{c}{Planning Steps} &
\multirow{2}{*}{Max Score} \\
 &  &  &
w/o Guidance & \textbf{w/ Guidance} &
w/o Guidance & \textbf{w/ Guidance} &
 \\
\hline
1   & 10 & 6.0 $\pm$ 1.8 & 7.0 $\pm$ 1.7 & \textbf{9.0 $\pm$ 0.0} & 18.1 $\pm$ 4.1 & \textbf{2.1 $\pm$ 0.3} & 9.0 \\
2 & 10  & 11.3 $\pm$ 1.4 & 13.6 $\pm$ 2.0 & \textbf{15.0 $\pm$ 0.0} & 12.4 $\pm$ 6.5 & \textbf{3.9 $\pm$ 2.0} & 15  \\
3      & 10 & 2.5 $\pm$ 0.4  & 2.2 $\pm$ 0.4 & \textbf{4.0 $\pm$ 0.0} & 20.0 $\pm$ 0.0 & \textbf{5.3 $\pm$ 2.9} & 4   \\
\hline
\end{tabular}
\end{table*}

\begin{figure*}[hbt!]
    \centering
    \includegraphics[width=\linewidth]{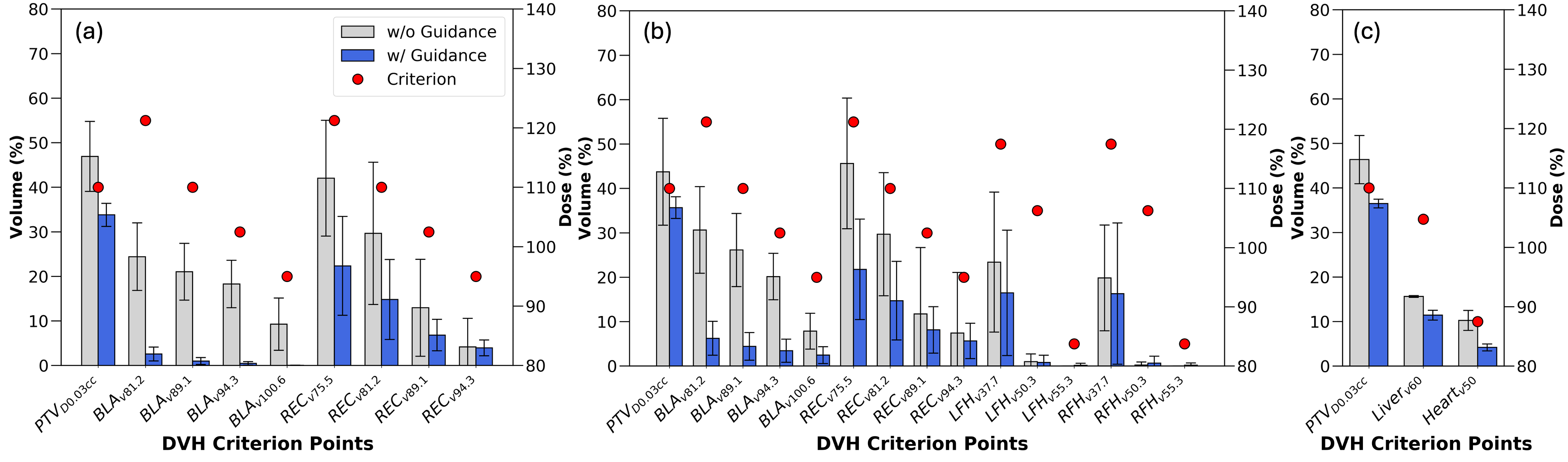}
    \caption{Quantitative comparison of dose-volume results for final treatment plans with and without DRL guidance across the three scenarios: (a) 7-beam prostate IMRT (2 OARs), (b) 180-beam prostate IMRT (4 OARs), and (c) 7-beam liver IMRT (2 OARs). Grey and blue distributions represent results from the LLM agent without and with DRL-derived TPP guidance, respectively. Red dots denote clinical evaluation criteria, and error bars indicate standard deviations. The right y-axis scales $PTV_{D0.03cc}$ dose (\%), while the left y-axis scales volume (\%) for all OAR criterion points.}
    \label{fig:doseVolumeStatistics}
\end{figure*}

To examine planning quality beyond the binary scoring system, Fig. \ref{fig:doseVolumeStatistics} presents the organ dose-volume distributions for the final plans. In general, the guided agent consistently achieves a lower mean and lower variance across all evaluation criteria compared to the unguided version. The maximum dose to the PTV ($PTV_{D0.03cc}$) is effectively regulated by the guided LLM agent, whereas unguided plans exhibit wide variance and exceed the 110\% dose threshold in 8/10, 4/10, and 10/10 cases for scenarios 1, 2, and 3, respectively. Regarding OAR sparing, the guided framework generates plans with much lower organ volumes than required under each dose evaluation points. In contrast, the unguided agent struggles to OAR sparing, particularly for rectum sparing ($REC_{V75.5}$ and $REC_{V81.2}$) in scenarios 1 and 2, and heart sparing ($Heart_{V50}$) in scenario 3. These results strongly suggest that when guided by TPP distributions derived from a small, site-specific cohort, an LLM planning agent can achieve optimal and consistent performance that generalizes across diverse treatment sites, beam arrangements, and organ configurations.

{}

\subsection{DRL-guidance Effects on LLM Planning Generality}
To better understand the source of the observed generality in the guided LLM planning agent, we performed a systematic investigation of the final TPP distributions and their correlation with the achieved dose-volume distributions.

\begin{figure*}[hbt!]
    \centering
 \includegraphics[width=1\linewidth]{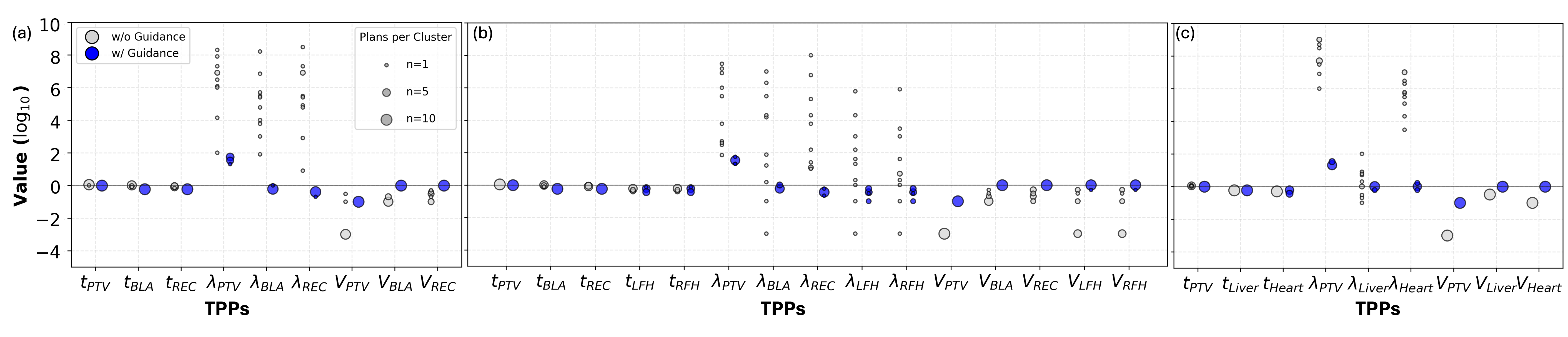}
    \caption{Distribution of TPP values for the final plans obtained by the LLM agent with and without guidance across the three planning scenarios: (a) scenario 1 (prostate IMRT with two OARs), (b) scenario 2 (prostate IMRT with four OARs), and (c) scenario 3 (liver IMRT). Marker size indicates the number of occurrences of a parameter value among the test cases, and the vertical axis shows the logarithmic scale of the parameter magnitude.}
    \label{fig:TPPSpacesCombined}
\end{figure*}

The distributions of final TPP configurations across the three scenarios are illustrated in Fig. \ref{fig:TPPSpacesCombined}. When comparing agents within the same scenario, the guided LLM agent produces highly concentrated and range-limited TPP values. In contrast, the unguided agent's parameters are significantly more scattered, spanning a much larger range. This is particularly evident in the $\lambda$ weight distributions, where guided values cluster near $10^2$ while unguided values frequently escalate to $10^9$. Across all three planning scenarios, both agents exhibit consistent, albeit different, TPP reasoning patterns. The guided agent demonstrates a clear hierarchical preference that closely mirrors the DRL-derived TPP distributions (Fig. \ref{fig:TPPSpaceGuidance}). Specifically, it maintains $t_\mathrm{PTV}$ near $10^0$ while setting OAR thresholds to lower values (between $10^{-1}$ and $10^0$). Similarly, it optimizes $V_\mathrm{PTV}$ to approximately $10^{-1}$ while keeping OAR volume parameters near $10^0$. Crucially, the agent performs targeted weight adjustments, raising $\lambda_\mathrm{PTV}$ to a magnitude of $10^1$ to $10^2$ while maintaining $\lambda_\mathrm{OARs}$ between $10^{-2}$ and $10^0$. This behavior indicates a successful implementation of M2M knowledge transfer. In comparison, the unguided agent shows similar directional preferences for $t$ and $V$ parameters but with much higher inter-case variance and value range (as low as $10^{-4}$ order). Its primary failure mode is a lack of discrimination over $\lambda$ parameters, where it tends to increase all $\lambda$ weights to extreme high values simultaneously. This behavior saturates the penally space, which fails to effectively trade off between the competing priorities of PTV coverage and OAR sparing and results in suoptimal plans.

It's worth mentioning that scenario 3 also provides a critical assessment of the agents' robustness against initialization bias, as the 10 planning cases were generated from a single patient anatomy using different TPP initializations. As illustrated in Fig. \ref{fig:TPPSpacesCombined}, the unguided agent proves highly sensitive to initial plan quality, resulting in widely divergent final TPP configurations. This lack of convergence leads to higher variance in final treatment plan quality, as seen in Fig. \ref{fig:doseVolumeStatistics} (c). In contrast, the guided agent demonstrates remarkable consistency, converging toward a similar TPP configuration regardless of the starting point. This stability significantly reduces planning quality variance, ensuring a reproducible, high-quality outcome even when starting from suboptimal initial parameters. 

{}

\begin{figure*}[hbt!]
    \centering
    \includegraphics[width=1\linewidth]{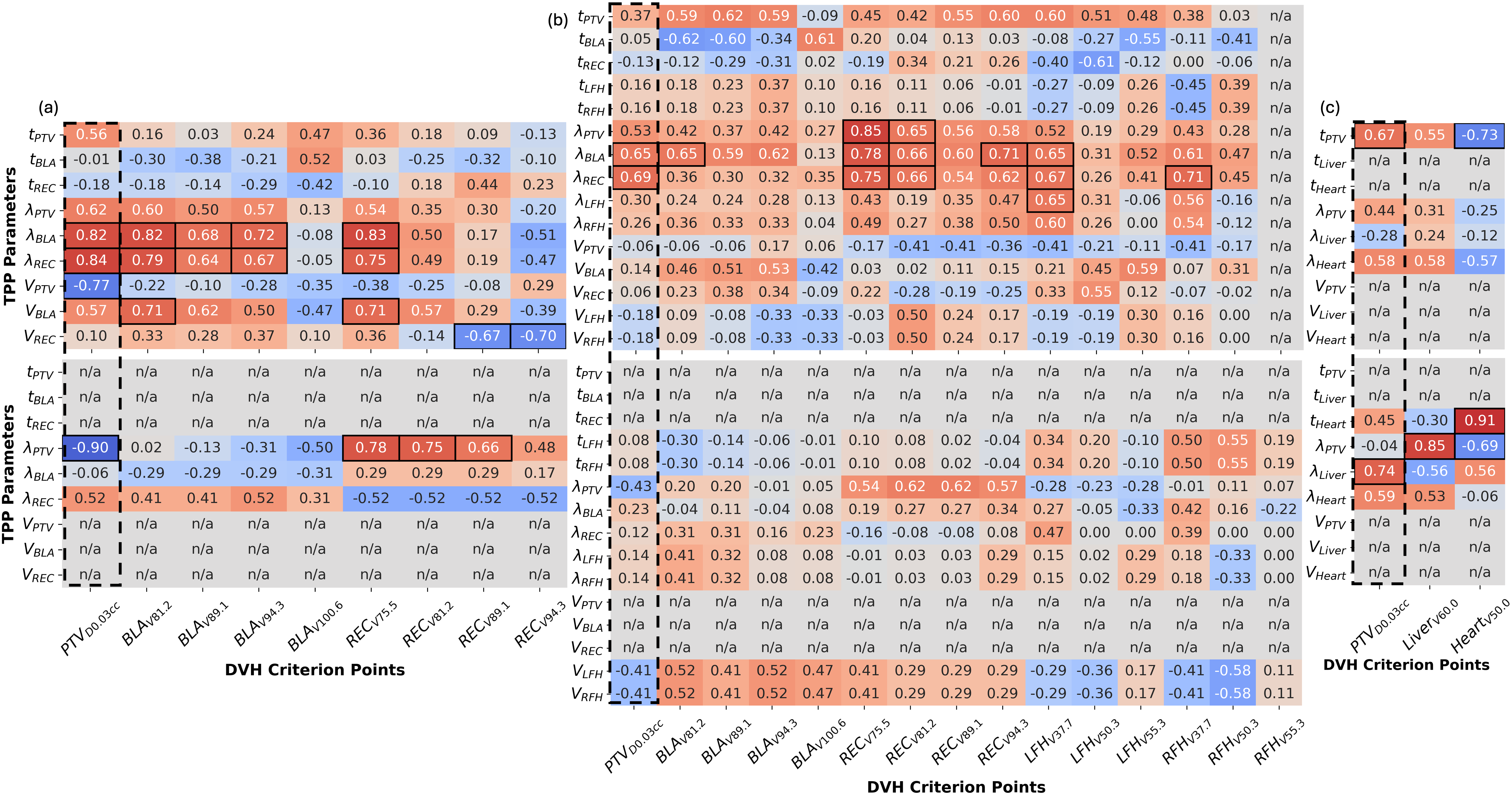}
    \caption{Spearman correlation analysis between treatment planning parameters (TPPs) and dose-volume metrics for plans with (bottom) and without (top) DRL guidance. Comparisons are shown for: (a) 7-beam prostate IMRT (2 OARs), (b) 180-beam prostate IMRT (4 OARs), and (c) 7-beam liver IMRT (2 OARs). Black boxes highlight significant correlations ($p < 0.05$). The 'n/a' designation indicates constant metrics across test cases where no valid correlation could be calculated.}
    \label{fig:tppdosecorrelation}
\end{figure*}

The regulatory effect of final TPP configurations on dose-volume distributions is further quantified via Spearman correlation analysis and illustrated in Fig. \ref{fig:tppdosecorrelation}. In the first column of each scenario, $\lambda_{\mathrm{PTV}}$ exhibits a negative correlation with $PTV_{D0.03cc}$ for the guided agent (reaching statistical significance in scenario 1), indicating that in the guided LLM, increasing the target weight effectively suppresses the PTV maximum dose. Conversely, the unguided agent exhibits a paradoxical positive correlation between $\lambda_{\mathrm{PTV}}$ and $PTV_{D0.03cc}$, a counter-intuitive result suggesting that without guidance, aggressive weight increases do not yield optimization gains and instead coincide with deteriorating target dose homogeneity. Furthermore, the effect of increasing $\lambda_{\mathrm{PTV}}$ on dose sparing of OARs is also clear in the guided agent (for instance, sparing bladder and sacrificing rectum in scenario 1), but no clear discrimination is found in the unguided agent. Similar patterns are observed in OAR regulation: for the guided agent, increases in $\lambda_{\mathrm{OAR}}$ are generally associated with a reduction in corresponding OAR volumes, maintaining a logical "cause-and-effect" relationship. In the unguided agent, this correlation often breaks down. Overall, these findings suggest that the M2M TPP guidance helps constrain the agent to a physically valid region of the TPP space and restores the optimization intuition required to ensure that the reasoning leads to predictable clinical improvements rather than paradoxical dose increases. Crucially, this physical region appears to be independent of the specific site, OAR configuration, and beam orientation.
{}
\section{Discussion}
In this study, we proposed an M2M knowledge-guided LLM planning framework for automated radiotherapy treatment planning. In the proposed framework, TPP distribution knowledge discovered by a DRL agent is transferred to an LLM agent through in-context learning, enabling the LLM to perform iterative treatment planning without human intervention. Experimental evaluations across three planning scenarios demonstrated that the DRL-guided LLM agent consistently achieved optimal planning scores while significantly reducing the number of planning iterations compared with unguided LLM planning. Beyond effectiveness and efficiency gains, the study of final TPP distributions revealed that the guided agent successfully learned appropriate TPP distribution ranges and their relative importance when navigating trade-offs between competing dose objectives. Under this meaningful guidance for decision-making, the agent effectively restores the logical "cause-and-effect" relationship between TPP tuning and dosimetric outcomes, which remains robust regardless of specific patient anatomy, treatment site, OAR configuration, or beam orientation. Furthermore, the results from scenario 3 demonstrated that the guided agent is insensitive to initial plan qualities, providing an additional level of planning robustness. Collectively, these findings indicate that the proposed framework is highly promising for achieving generalizable and trustworthy autonomous treatment planning.

Despite these encouraging results, several limitations should be noted. First, while the selected planning scenarios provided a robust initial test of the proposed framework, they remain relatively simplified. Although scenario 2 introduced increased complexity on OAR and beam configurations, the overall configurations may not yet capture the full scale of complexity in realistic clinical cases. Therefore, future studies are required to evaluate the framework under increasingly realistic and complex treatment planning scenarios. Our immediate next step involves applying this prototype framework to standard clinical planning settings to validate its real-world utility.

Second, the reasoning performance of LLMs is highly sensitive to prompt engineering. Prior research has demonstrated that advanced prompting strategies, such as chain-of-thought (CoT) reasoning, can significantly enhance LLM performance in radiotherapy by explicitly structuring intermediate decision-making steps \cite{yang2025zero}. However, the efficacy of such strategies remains limited in fully autonomous treatment planning settings where human-provided parameter ranges are still required to bound the TPP search space \cite{yang2025zero}. While the present study focuses on establishing the feasibility and effectiveness of an M2M knowledge transfer paradigm that eliminates this need for human intervention, future iterations could incorporate structured reasoning frameworks like CoT. Integrating CoT prompts with DRL-derived guidance may further strengthen the logical consistency, planning precision, and overall robustness of the LLM agent when navigating highly complex clinical environments.

Nevertheless, several insights can be derived from this study. The proposed M2M knowledge transfer paradigm demonstrates a promising strategy for combining the specialized optimization of reinforcement learning with the versatile reasoning of LLMs in complex decision-making tasks. As shown in section \ref{sec: results}, while an unguided LLM agent can iteratively adjust parameters, it often fails to converge due to an inability to distinguish between competing objectives, frequently resulting in extreme, saturated parameter values. In contrast, the guided LLM agent demonstrates a logical reasoning process that improves plan quality by mirroring the DRL agent’s expertise within a physically valid TPP distribution. While DRL enables deep exploration of the solution space to achieve superhuman performance in specific domains, the LLM provides the flexible adaptation required to apply that knowledge across different task settings. By successfully transferring knowledge from a DRL agent trained on prostate cases to liver and multi-OAR scenarios, this framework demonstrates the potential to achieve both high-quality solutions and broad clinical flexibility.

Furthermore, the analysis of final TPP distributions and Spearman correlations suggests that optimal treatment plans exhibit identifiable, site-independent parameter patterns. These patterns reflect underlying planning principles, specifically a hierarchical priority of targets over OARs, that restore a logical "cause-and-effect" relationship to the tuning process. From an inverse optimization perspective, planning problems across diverse sites share similar mathematical structures; consequently, the TPP configurations that yield clinical success could exhibit common topological features. This observation suggests that the TPP space discovered via DRL constitutes a "physical effective zone" for optimization. A further in-depth investigation over the TPP space for treatment planning in realistic clinical settings holds the potential toward a truly generalizable automated treatment planning system that remains effective regardless of specific patient anatomy or beam configuration.

\section{Conclusion}
In conclusion, this work proposes an M2M knowledge-guided LLM framework for automated radiotherapy treatment planning. By transferring TPP space knowledge discovered by a DRL agent to an LLM agent through in-context learning, the framework enables efficient and effective planning automation. These results highlight the potential of M2M knowledge transfer for generalizable automated treatment planning systems.

\section*{References}
\bibliographystyle{IEEEtran}
\bibliography{references}

\end{document}